# The flavin reductase ActVB from *Streptomyces coelicolor*. Characterization of the electron transferase activity of the flavoprotein form.

Laurent Filisetti, Julien Valton, Marc Fontecave, and Vincent Nivière*

Laboratoire de Chimie et Biochimie des Centres Redox Biologiques, DRDC-CEA/CNRS/Université Joseph Fourier, 17 Avenue des Martyrs, 38054 Grenoble, Cedex 9, France.

* To whom correspondence should be addressed. Vincent Nivière; Tel.: 33-4-38-78-91-09; Fax: 33-4-38-78-91-24; E-mail: vniviere@cea.fr.

ABSTRACT

The flavin reductase ActVB is involved in the last step of actinorhodin biosynthesis in *Streptomyces coelicolor*. Although ActVB can be isolated with some FMN bound, this form was not involved in the flavin reductase activity. By studying the ferric reductase activity of ActVB, we show that its FMN-bound form exhibits a proper enzymatic activity of reduction of iron complexes by NADH. This shows that ActVB active site exhibits a dual property with regard to the FMN. It can use it as a substrate that goes in and off the active site or as a cofactor to provide an electron transferase activity to the polypeptide.

INTRODUCTION

Flavin reductases represent a broad class of enzymes defined by their ability to catalyze the reduction of free flavins, riboflavin, FMN or FAD, by NAD(P)H [1-4]. A clear distinction between a flavin reductase and a flavoprotein resides in the fact that the former uses flavin as a substrate whereas the latter uses it as a prosthetic group. In general, flavin reductase physiological functions have been directly associated with the chemical properties of their released product, the free reduced flavin [1]. Free reduced flavins are known to reduce natural iron complexes like ferrisiderophores very efficiently and thus flavin reductases have been proposed to play an important role in iron uptake in bacteria, as ferric reductase enzymes [1]. Flavin reductases have been also found to be essential for a recently emerging class of flavin dependent monooxygenases such as luciferase for light emission or enzymes involved in antibiotics synthesis, aromatic and sulfur oxidations and other reactions [2-4]. In those systems, the reduced free flavin generated by the flavin reductase binds to the active site of the monooxygenase and, in the presence of oxygen, is then converted into a flavinhydroperoxide species as the key intermediate for oxidation reactions [5]. Such a two-component system has been recently studied in the case of the last step of actinorhodin biosynthesis, a natural antibiotic produced by *Streptomyces coelicolor* [2-3, 5]. It involves ActVB, as the flavin reductase [2,3], and ActVA-ORF5, as the monooxygenase component [5]. Although ActVB can be isolated with various amount of FMN bound, depending on the preparation, we unambiguously demonstrated that flavin reductase activity of ActVB did not involve the protein-bound FMN. The presence of protein-bound FMN was rather interpreted as a reflect of the strong affinity of ActVB for the oxidized FMN [3].

In this work, by studying the ferric reductase activity of ActVB, we show that its FMN-bound form exhibits a proper enzymatic activity namely the reduction of iron complexes by NADH, in the absence of additional free flavins. These data highlight a dual property of ActVB active site with regard to the FMN. In the presence of high concentrations of FMN, it can use it as a substrate that goes in and off the active site. In the absence of free flavin, the ActVB-bound FMN provides the polypeptide a flavoprotein character with an electron transferase activity.

MATERIALS AND METHODS

NADH, FMN, $Fe^{3+}$-EDTA, ferricyanide and bathophenanthroline were from Sigma. Pyoverdin, azotobactin and desferal were gifts from Dr. Isabelle Schalk (University of Strasbourg, France). The recombinant ActVB protein from *Streptomyces coelicolor* was purified as reported in [3]. ActVB FMN content was determined spectrophotometrically using a $\varepsilon_{455nm} = 13.6$ mM$^{-1}$ cm$^{-1}$ [3]. Enzymatic activities were measured at 18 °C, under anaerobiosis conditions. Flavin and $Fe^{3+}$-EDTA reductase activities were determined from the decrease of the absorbance at 340 nm ($\varepsilon_{340nm}= 6.22$ mM$^{-1}$ cm$^{-1}$), as reported in [3]. For the low redox potential iron complexes, azotobactin, pyoverdin and desferal, the reduction reaction was carried out in the presence of the ferrous iron acceptor bathophenanthroline and followed from the increase of the absorbance at 535 nm ($\varepsilon_{535nm}= 25.0$ mM$^{-1}$ cm$^{-1}$), characteristic of the $Fe^{2+}$-bathophenanthroline complex [6]. Ferricyanide reduction was followed from the decrease of the absorbance at 420 nm ($\varepsilon_{420nm}= 1.01$ mM$^{-1}$ cm$^{-1}$). Anaerobic experiments were carried out in a Jacomex glove box equipped with an Uvikon XL spectrophotometer coupled to the measurement cell by optical fibers (Photonetics system).

RESULTS

*Fe-EDTA reductase activity of ActVB in the presence of free flavins*

The ability of ActVB to reduce $Fe^{3+}$-EDTA complex in the presence of NADH and FMN was tested in anaerobiosis. The reaction was followed spectrophotometrically from the decay of the absorbance at 340 nm, due to the oxidation of NADH. As shown in Figure 1, the extent of NADH oxidation was found directly proportional to the amount of $Fe^{3+}$-EDTA present in the reaction mixture, giving a proportion of one NADH molecule reducing two molecules of $Fe^{3+}$-EDTA (Figure 1). In the absence of $Fe^{3+}$-EDTA, only a slight amount of NADH was oxidized, corresponding to the amount of free FMN present in the assay mix, (5 µM, data not shown). When the initial velocity of the reaction was studied in the presence of a fixed concentration of $Fe^{3+}$-EDTA (300 µM) as a function of NADH or FMN concentrations, typical Michaelis-Menten curves were obtained (data not shown). $V_m$ and $K_m$ values for NADH and FMN under those conditions were found to be identical to those reported using the standard flavin reductase activity assay in which the electron acceptor is oxygen (Table I and [3]). These data suggest that in the presence of free FMN, reduction of $Fe^{3+}$-EDTA occurs through a chemical reaction with the reduced FMN, provided by the flavin reductase activity of ActVB.

*Fe-EDTA reductase activity of ActVB in the absence of free flavins*

The ActVB polypeptide contains various amounts of bound FMN, depending on the enzyme preparation [3]. Reduction of $Fe^{3+}$-EDTA by NADH catalyzed by different ActVB preparations containing 0.1 or 0.5 mol of bound–FMN per mol of polypeptide chain was tested in the absence of free FMN. In anaerobiosis, in the presence of 300 µM $Fe^{3+}$-EDTA, the velocity of NADH oxidation was found to be directly proportional to the total concentration of FMN bound to ActVB (Figure 2). It should be noted that under these conditions, the concentration of FMN bound to ActVB is several orders of magnitude smaller than the concentration of free flavins used in the flavin reductase activity assay described above (Figure 1). These data show that the ActVB-FMN complex catalyzes the reduction of $Fe^{3+}$-EDTA by NADH in the absence of free flavin. From these data, a $k_{cat}$ value of 17 $s^{-1}$ for the ActVB protein containing one mol of bound-FMN per mol of polypeptide chain can be calculated.

The mechanism for the enzyme reaction was investigated by studying the initial velocity of NADH oxidation as a function of NADH and $Fe^{3+}$-EDTA concentrations, under anaerobiosis conditions. As shown in Figure 3A, the double reciprocal plot of the initial velocity of NADH oxidation as a function of NADH concentration at several levels of $Fe^{3+}$-EDTA shows a series of parallel lines. In Figure 3B, the initial velocity of the reaction as a function of $Fe^{3+}$-EDTA concentration at several levels of NADH also shows a series of parallel lines. These data are consistent with a ping-pong enzyme mechanism. The kinetic parameters, $K_m$ and $V_m$ values determined from Figure 3 are shown in Table I. The $k_{cat}$ value of 9 $s^{-1}$ corresponding to an ActVB protein containing 0.5 mol of bound-FMN per mol of polypeptide chain is in full agreement with the value previously determined for the ActVB protein containing one mol of FMN per mol of polypeptide chain.

That the reduced FMN bound to ActVB can indeed transfer electrons to the $Fe^{3+}$-EDTA, according to the ping-pong mechanism, was verified with the following experiments. The FMN bound to ActVB was reduced anaerobically with one molar equivalent of NADH, with respect to FMN (Figure 4). Complete reduction was obtained as shown by the disappearance of the 457 nm band characteristic for the oxidized FMN bound to ActVB [3], together with the appearance of a $NAD^+$-$FMN_{red}$ charge transfer band in the 550 to 800 nm region, with an isosbestic point at 507 nm [3]. Additions of $Fe^{3+}$-EDTA resulted in an immediate oxidation of the FMN, as shown by the increase of the absorption at 457 nm, together with the decrease of the charge transfer band, again with an isobestic point at 507 nm (Figure 4). Complete oxidation was achieved when 2 molar equivalents of $Fe^{3+}$-EDTA per mol of reduced FMN in the active site of ActVB was added (Inset of Figure 4). No evidence for the appearance of half-reduced flavin forms could be observed during the oxidation process. The resulting fully oxidized FMN exhibits a maximal absorption at 457 nm, which indicates that at the end of the oxidation process the flavin remains in the active site of ActVB [3].

*Ferric reductase activity using various ferric complexes*

Different ferric complexes, including ferrisiderophores, were tested for their ability to be substrates for the ferric reductase activity of the ActVB-FMN complex. As shown in Table I, the ActVB-FMN complex did not catalyze reduction of ferrioxamine. On the contrary, ferricyanide, ferripyoverdin and ferriazotobactin were found to be substrates for the enzyme, with $K_m$ values comparable to that reported for the $Fe^{3+}$-EDTA complex (Table I). On the other hand, $V_m$ ($k_{cat}$) values varied, with ferricyanide as the best substrate and ferripyoverdin as the less efficient (Table I).

DISCUSSION

Although the as-isolated ActVB is a mixture of FMN-free and FMN-containing forms, our previous work demonstrated that the bound FMN was not involved in the flavin reductase activity of ActVB [3]. ActVB was shown to contain only one FMN binding site, which was then defined as the substrate-binding site for the flavin reductase activity. So that ActVB was not initially described as a flavoprotein [3].

In the present work, by exploring the capability of ActVB to catalyze the reduction of several iron complexes by NADH, we demonstrate that it can also behave as a functional flavoprotein, with a bound FMN cofactor that transfers electrons from NADH to the iron complex. These data highlight a new functionality of ActVB, which thus can use FMN not only as a substrate but also as a cofactor.

When a ferric complex was tested for reduction in the presence of free flavin, ActVB exhibited a classical ferric reductase activity through its flavin reductase activity [1]. This is due to its capacity to bind and reduce free flavins at the expense of NADH and release the reduced flavins that transfer electrons to $Fe^{3+}$, non-enzymatically. [1].

On the other hand, ActVB can also catalyze the reduction of various iron complexes in the absence of free flavin and this ferric reductase activity has been shown to directly depend on the amount of FMN bound to the ActVB polypeptide. Under these conditions, the concentration of FMN in the activity test is several orders of magnitude smaller than the $K_m$ value for FMN related to the flavin reductase activity. Therefore, it can be excluded that this specific ferric reductase activity results from the flavin reductase activity of ActVB.

The presence of a redox cofactor in ActVB is expected to induce a specific enzyme mechanism, with a transfer of the reducing equivalents from NADH to the iron complexes through the FMN cofactor. Indeed we show that when reduced by NADH, the FMN cofactor can directly reduce the ferric complex. Also, the double reciprocal plots of the initial velocity of the ferric reductase activity shows a series of parallel lines when one of the two substrates, NADH or the iron complex, varied. This suggests a two-step transfer ping-pong enzyme mechanism, as it has been proposed for the flavoprotein spinach ferredoxin-NADP(H) reductase for the reduction of ferredoxin in the presence of NADPH [7]. For the ferric reductase activity of ActVB, such a mechanism is consistent with NADH reducing first the FMN cofactor, which in turn reduces two molecules of iron complexes:

$$\text{ActVB-FMN} + \text{NADH} \leftrightarrow \text{ActVB-FMN-NADH}$$

$$\text{ActVB-FMN-NADH} \leftrightarrow \text{ActVB-FMNH}_2 + \text{NAD}^+$$

$$\text{ActVB-FMNH}_2 + \text{Fe}^{3+}(\text{EDTA}) \leftrightarrow \text{ActVB-FMNH}^\bullet + \text{Fe}^{2+}(\text{EDTA})$$

$$\text{ActVB-FMNH}^\bullet + \text{Fe}^{3+}(\text{EDTA}) \leftrightarrow \text{ActVB-FMN} + \text{Fe}^{2+}(\text{EDTA})$$

Based on a three-dimensional crystal structure, a similar enzyme mechanism was proposed for FeR, a ferric reductase from *Archaeoglobus fulgidus* that exhibits some sequence homologies with ActVB [8]. FeR was believed to contain an FMN cofactor so far [8]. However, in the case of FeR, no kinetic data were reported to support such a mechanism [8].

It should be noted that in the case of ActVB, the formation of an intense charge transfer band between $NAD^+$ and $FMNH_2$ within the active site [3] might not be totally in agreement with such a ping-pong mechanism where $NAD^+$ has to dissociate from the charge transfer complex before the iron complex get reduced. An ordered mechanism with the formation of ternary complex between the FMN cofactor, NADH and the iron complexes previous to the electron transfer through the FMN cofactor might be also considered [7]. In the case of the flavoprotein ferredoxin reductase, such a mechanism was proposed to be also consistent with a series of parallel lines [7], as observed in this work. Further studies are needed to clarify this point.

In addition to $Fe^{3+}$-EDTA, FMN-containing ActVB can efficiently catalyze the reduction of several iron complexes and ferrisiderophores, which display comparable $K_m$ values. Because these molecules exhibit different structures, this clearly indicates that $K_m$ values do not just reflect the affinity for the ferric complexes substrates, but rather a complex combination of rate constants of individual steps of the two-substrate enzyme mechanism [9].

Thus using ferric complexes as substrates, we demonstrate here that, in the absence of free flavins, ActVB displays an efficient electron-transferase activity. This might have physiological consequences with regard to iron metabolism in *S. coelicolor*, but also to other metabolic pathways, where ActVB would used to reduce other electron-accepting substrates.

In conclusion, these studies on the ferric reductase activity of ActVB reveal that ActVB exhibits a dual property with regard to the FMN molecule. It can use it as a substrate that goes in and off the active site or as a cofactor that confers to the polypeptide an electron transferase activity. To our knowledge, this is the first time that such a property has been described for a polypeptide chain.

These data imply that the same polypeptide chain has different affinities for the flavin molecule. This is difficult to explain if one consider a single flavin binding site on the ActVB polypeptide, as illustrated by the crystal structure of proteins from the same family, FeR [8] and PheA2 [10], in complex with a flavin. Whether the ActVB homodimer would enjoy cooperativity properties as far as FMN binding is concerned, with strong binding of the first flavin (cofactor) and weaker binding of the second flavin (substrate), remains to be investigated. Interestingly, in FeR, which is a homodimer like ActVB, only one subunit has been observed to bind a FMN cofactor [8]. Characterization of two binding sites on the ActVB homodimer, one per monomer with different affinities for FMN, is currently under investigation.

ACKNOWLEDGEMENTS

This work has been supported by an Emergence Région Rhône-Alpes Fellowship. We thank Dr. Isabelle Schalk (Strasbourg, France) for the generous gift of ferrisiderophore molecules.

FIGURE LEGENDS

Figure 1. Time course of NADH (200 μM) oxidation followed at 340 nm in anaerobiosis in the presence of 3 μM FMN, 0.054 μM ActVB and different concentrations of $Fe^{3+}$-EDTA: 15, 30, 60, 100 and 350 μM, in 50 mM Tris/HCl pH 7.6.

Figure 2. Ferric reductase activity in the absence of free FMN, followed at 340 nm in anaerobiosis. The solution contains 200 μM NADH, 300 μM $Fe^{3+}$-EDTA in 50 mM Tris/HCl pH 7.6. The reaction was started by addition of various preparations of ActVB-bound FMN protein. (◆) ActVB 0.054 μM, FMN-bound 0.005 μM, (▲) ActVB 0.490 μM, FMN-bound 0.05 μM, (□) ActVB 0.110 μM, FMN-bound 0.055 μM, (●) ActVB 0.327 μM, FMN-bound 0.16 μM. In the inset is shown the initial velocity of NADH oxidation as a function of the concentration of the ActVB-bound FMN.

Figure 3. Ferric reductase initial velocity for the ActVB FMN bound enzyme. The reaction was followed at 340 nm in anaerobiosis, in 50 mM Tris/HCl pH 7.6. The ActVB concentration used was 0.054 μM and the protein-bound FMN was 0.027 μM. A, as a function of NADH concentration, in the presence of (◆) 16, (▲ 20, (□) 33, (●) 65 and (Δ) 110 μM $Fe^{3+}$-EDTA. B, as a function of $Fe^{3+}$-EDTA concentration, in the presence of (◆) 1.3, (▲) 2.3, (□) 3.9, (●) 8.6 and (Δ) 16 μM NADH.

Figure 4. Oxidation of reduced ActVB by $Fe^{3+}$-EDTA. ActVB (150 μM) containing 80 μM protein-bound FMN was reduced in anaerobiosis with 80 μM NADH (●). Then, 30 (■), 60 (▲), 90 (○), 120 (□) and 150 (Δ) μM Fe-EDTA, final concentrations, were added. The inset shows the variation of the absorbance at 457 nm as a function of the [Fe-EDTA]/[FMN] ratio.

Table I. Kinetic parameters for the ferric reductase activity of ActVB[a]

|  | In the presence of free FMN[b] | | In the absence of free FMN | | | | | |
|---|---|---|---|---|---|---|---|---|
| Varied substrate | NADH | FMN | NADH[c] | Fe-EDTA[c] | Ferricyanide[d] | Pyoverdin[d,e] | Azotobactin[d,e] | Desferal[d,e] |
| $k_{cat}$ (s$^{-1}$) | 9.2±0.4 | 9.2±0.4 | 8.9±0.5 | 9.8±0.4 | 42±2 | 0.23±0.03 | 3.7±0.4 | No activity |
| $K_m$ (µM) | 6.6±0.5 | 1.0±0.1 | 8.4±0.5 | 105±14 | 130±20 | 128±38 | 165±40 | / |

[a]Measured in anaerobiosis in 50 mM Tris/HCl pH 7.6. The ActVB preparation contains 0.5 mol of FMN bound per mol of polypeptide chain. [b]With a fixed concentration of Fe$^{3+}$-EDTA (300 µM). [c]Data from Figures 3A and 3B. [d] With a fixed concentration of NADH (200 µM). [e]In the presence of 2 mM bathophenanthroline

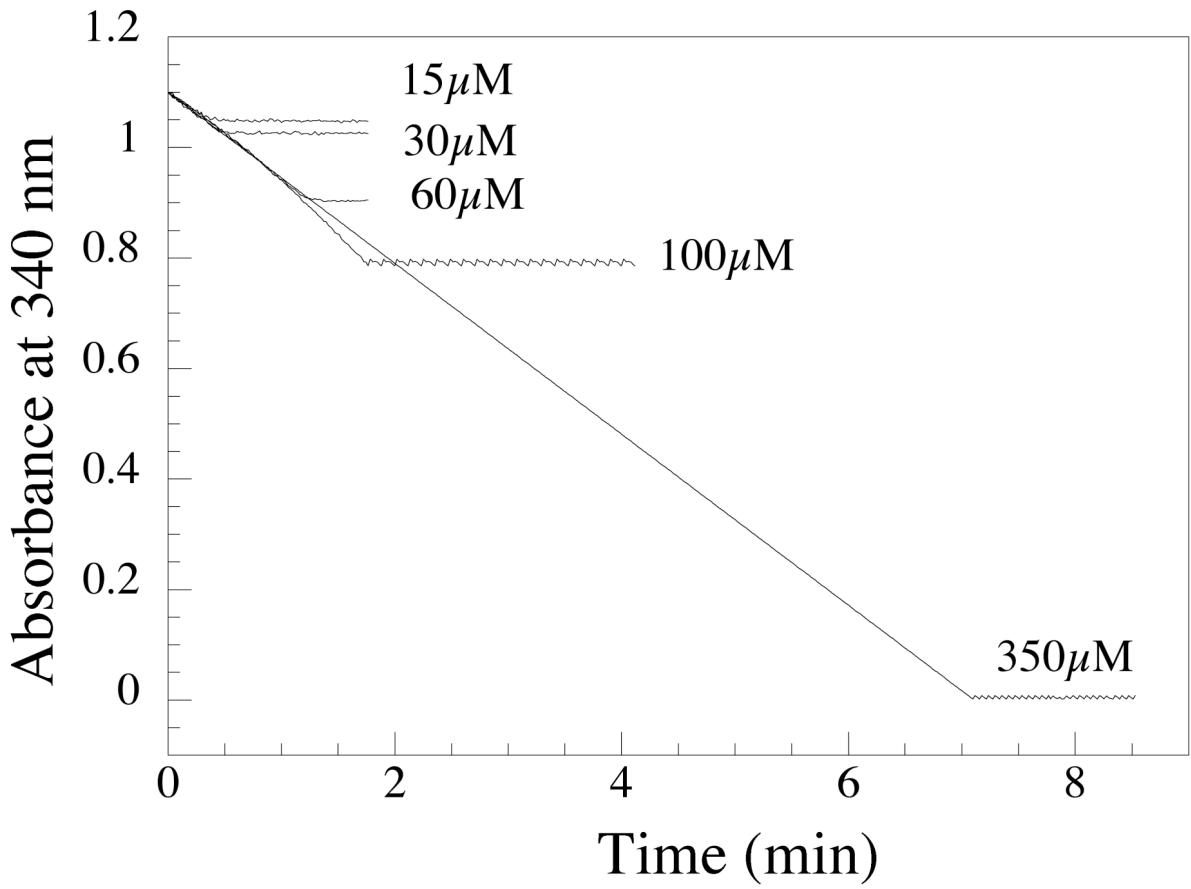

Fig. 1

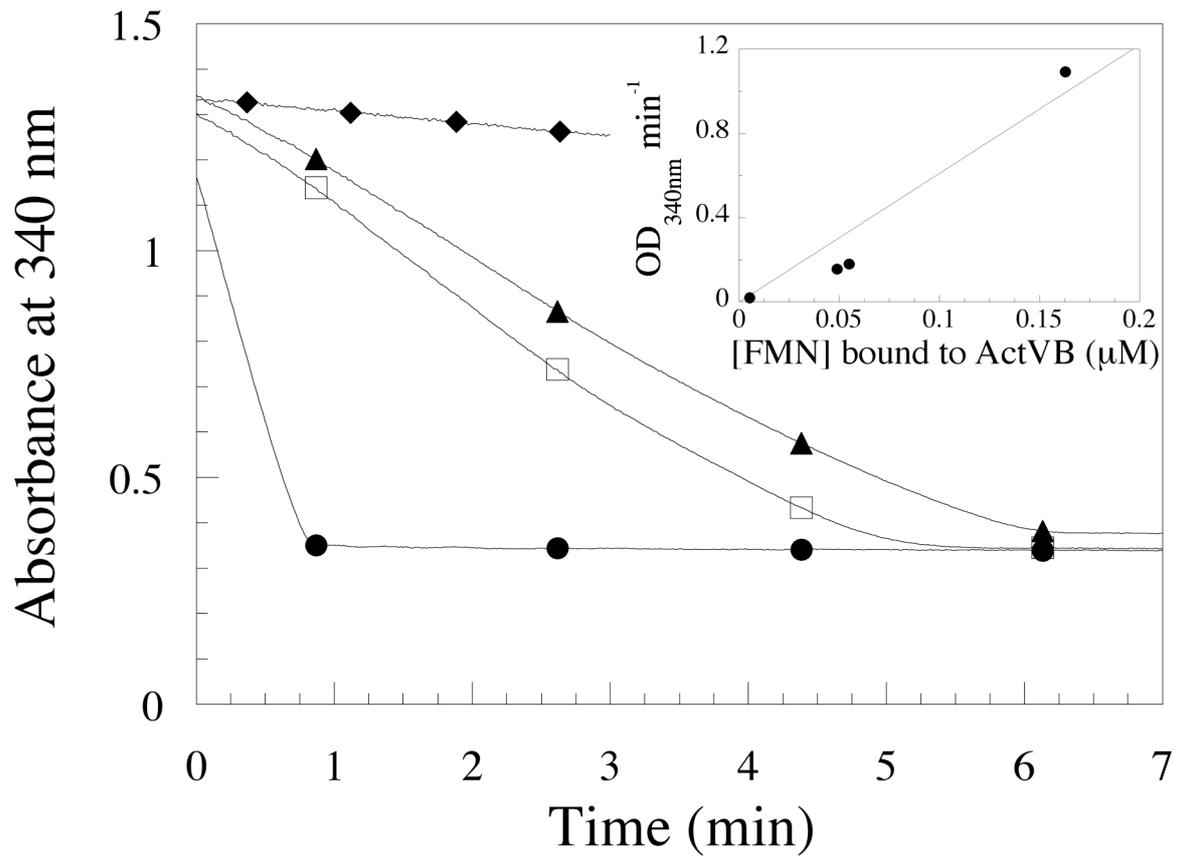

Fig. 2

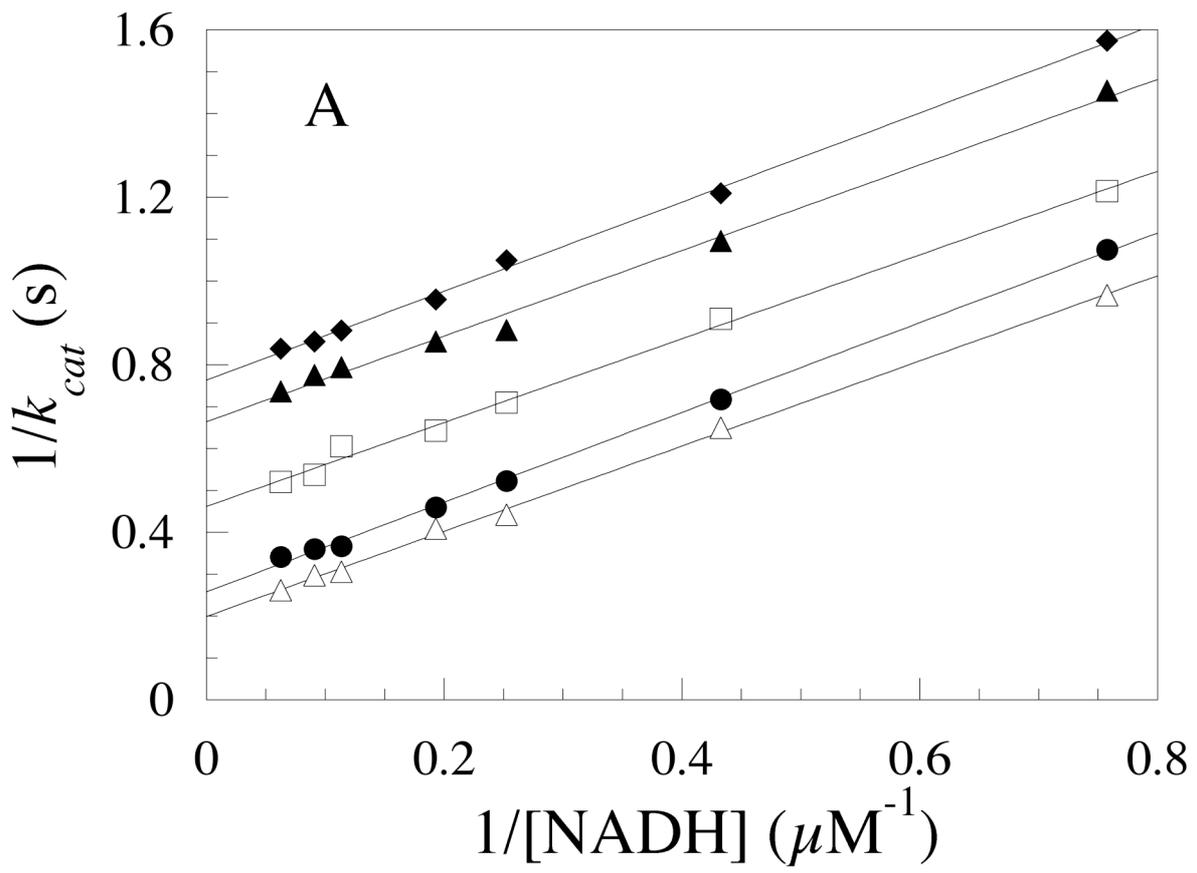

Fig. 3A

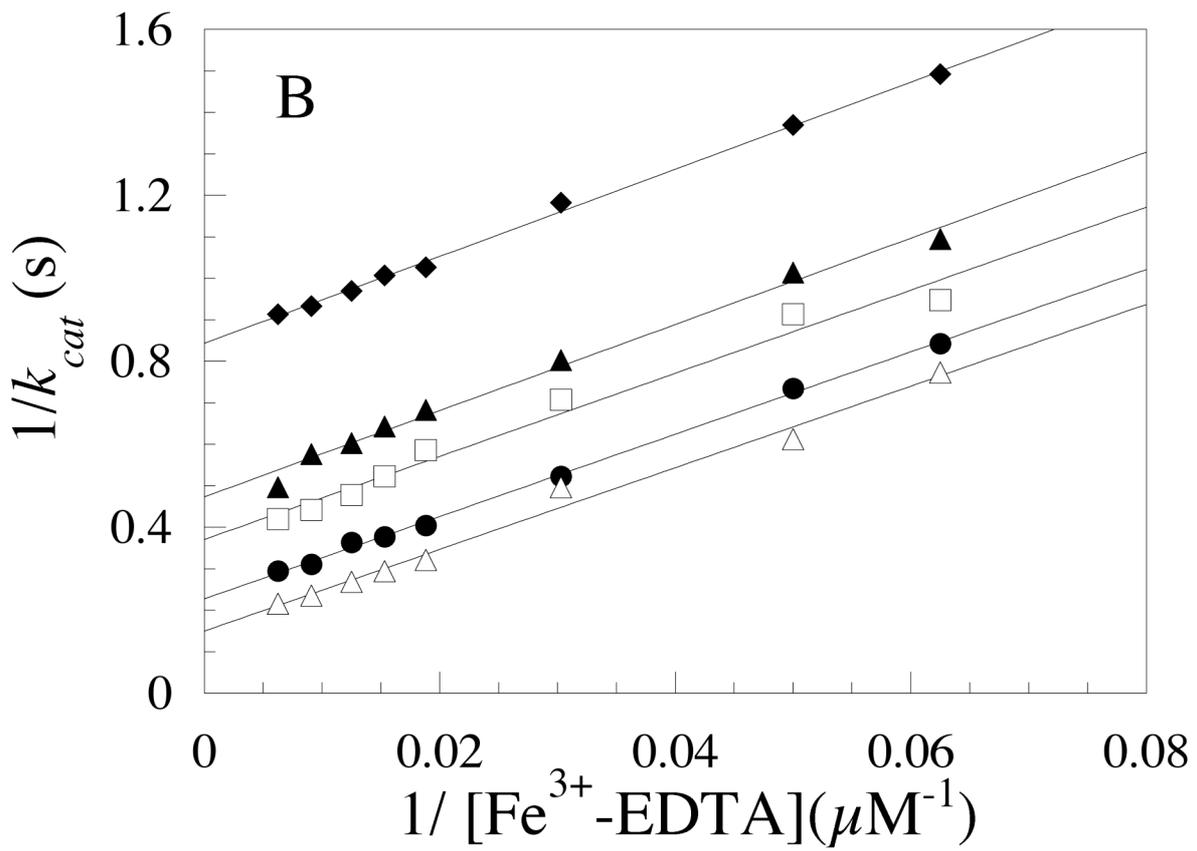

Fig. 3B

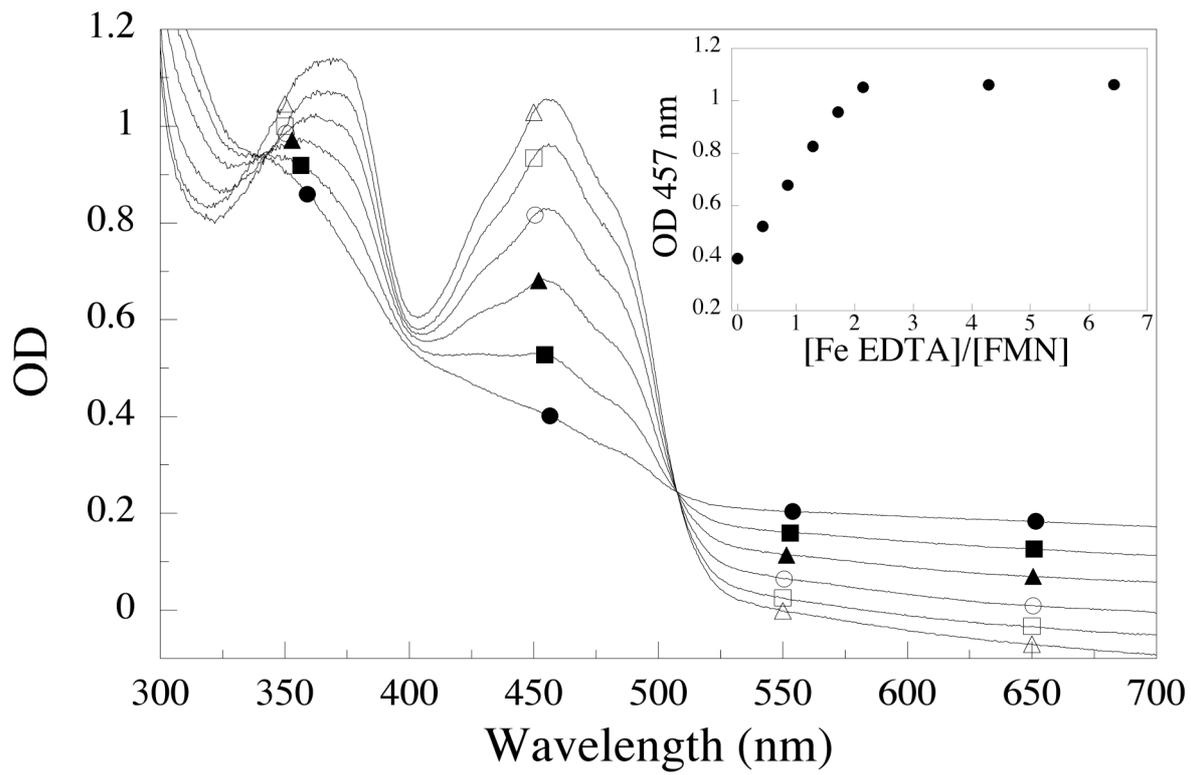

Fig. 4